\documentclass[aps,preprint]{revtex4}%
\usepackage{amsfonts}
\usepackage{amsmath}
\usepackage{amssymb}
\usepackage{graphicx}%
\providecommand{\U}[1]{\protect\rule{.1in}{.1in}}

\begin{document}
\title{Quintessence and tachyon dark energy in interaction with dark matter:
Observational constraints and model selection}
\author{Sandro M. R. Micheletti}
\email{smrm@fma.if.usp.br}
\affiliation{Universidade Federal do Rio de Janeiro, Campus Maca\'{e}, Avenida Alu\'{\i}zio
da Silva Gomes 50, Granja dos Cavaleiros, 27930-560, Maca\'{e}, Rio de
Janeiro, Brazil}

\begin{abstract}
We derive two field theory models of interacting dark energy, one in which
dark energy is associated with the quintessence and another in which it is
associated with the tachyon. In both, instead of choosing arbitrarily the
potential of scalar fields, these are specified implicitly by imposing that
the dark energy fields must behave as the new agegraphic dark energy. The
resulting models are compared with the Pantheon supernovae sample, CMB
distance information from Planck 2015 data, baryonic acoustic oscillations
(BAO) and Hubble parameter data. For comparison, the noninteracting case and
the $\Lambda CDM$ model also are considered. By use of the \textit{AIC} and
\textit{BIC} criteria, we have obtained strong evidence in favor of the two
interacting models, and the coupling constants are nonvanishing at more than
$3\sigma$ confidence level.

\end{abstract}
\maketitle


\thispagestyle{empty}

\section{Introduction}

Twenty years ago, two groups have discovered independently that the universe is in a period of accelerated expansion \cite{accelerated1} \cite{accelerated2}. In order to obtain such an acceleration in the expansion rate in the framework of the General Relativity, it is necessary that the universe be dominated by a component with negative pressure. Such component was called dark energy, and the first candidate considered for the dark energy was the cosmological constant. In fact, the universe model based on a cosmological constant and cold dark matter, the $\Lambda CDM$, has been capable of reproducing all observational data until now. However, there are two well-known problems with the cosmological constant: the fine tunning and the coincidence problems. The first of these problems arises when one try to associate the dark energy with the vacuum energy. Order-of-magnitude estimate from quantum field theory furnishes a value for the vacuum energy density about 120 orders of magnitude larger than the observed value of the dark energy density. This huge difference can be canceled by introducing counter terms, but these must be accurate to 120 decimal places, which is very unlikely to happen. On the other hand, the matter energy density is the same order of magnitude of the dark energy density today, but matter energy density scales as $a^{-3}$, where $a$ is the scale factor, so that in the past it was different from the energy density of the cosmological constant by many orders of magnitude. So, arises the question: why just now they are of the same order of magnitude? In order to solve such problems, many alternatives to the cosmological constant have been proposed.

The small value of the dark energy density as compared with the theoretical estimate for the vacuum energy density has led the idea that the dark energy is associated with a light scalar field rolling to the minimum of a self-interaction potential,
in such a manner that the field energy density decreases with the expansion of the universe more slowly than matter energy density. Rolling scalar fields have been extensively studied as alternatives to the cosmological constant, see e.g. \cite{peebles} - \cite{reviewLi}.
Considerable work also has been done on phenomenological models, where the pressure to energy density ratio of the dark energy - the equation of state parameter $\omega_{DE}\equiv
\frac{P_{DE}}{\rho_{DE}}$ - is an arbitrary function of the redshift, see e.g. \cite{copeland} and \cite{reviewLi}. Recently, tensions between the $\Lambda CDM$ and the observational data have appeared \cite{PLANCK2015} \cite{DES}, and models of modified gravity can alleviate some of these tensions, with an improvement in
the goodness of fit \cite{PLANCK2015} \cite{PLANCK2018}. Modified gravity theories as alternatives to obtain accelerated expansion also have been extensively studied, see e.g. \cite{copeland} and \cite{reviewLi}. Models which
include an interaction between dark energy and dark matter also can alleviate some of these tensions \cite{PLANCK2015}, \cite{interacaoelcio} - \cite{tensaoH0}. An advantage of the interaction is that dark energy and dark matter will evolve in a coupled fashion, and if dark energy decays into dark matter, this will at least alleviate the coincidence problem, the second of the problems with the cosmological constant mentioned above. Many papers considering an interaction between dark energy and dark matter have been published, and some evidence of the interaction has been found, see e.g. \cite{interacaoelcio}, \cite{sandro} - \cite{interacaoevidencia}. For a more complete list of references about evidences of the interaction, as well as for a discussion of theoretical aspects and cosmological implications, see \cite{revinteracaoelcio}. However, in the most of these papers, the interaction term in the equations of the model is derived phenomenologically. Much smaller is the number of papers where the interaction term is derived from a field theory. Examples of works in this direction are shown in \cite{Sandro3}, \cite{bean} - \cite{Sandro2}. In fact, if we suppose that the dark energy is associated to a physical field, it is more natural to assume that dark energy interacts with dark matter, as fields generally interact, unless such an interaction be prevented by some specific symmetry. In this paper, we will follow this path.

It is very common to choose scalar fields as candidates to dark energy, as the canonical scalar field, called quintessence, or the tachyon field. They naturally arise in particle physics and string theory. Scalar fields are also introduced in theories of
inflation. For reviews about the use of scalar fields as dark energy, see e.g. \cite{copeland}, \cite{reviewLi} and \cite{Sami}. The quintessence has the equation of state parameter, $\omega_{q}$, between $-1$ and $+1$. Quintessence models were investigated, e.g. in \cite{bean}, \cite{canonico} - \cite{canonico6}. The tachyon field has the equation of state parameter  $-1\leq\omega_{t}\leq0$. The tachyon Lagrangian
was derived from brane developments in string theory \cite{sen} - \cite{sen6}.
Tachyon as dark energy was studied, e.g. in \cite{Sandro3}, \cite{taquions} - \cite{taqholo2}. A natural question which arises is to choose the
potential $V\left(  \varphi\right)  $ of the scalar field. Common choices are power law or exponential potentials. However, these choices are in fact arbitrary. It would be interesting to
choose the potential by some physical criterion. Efforts in this direction were made in \cite{Sandro1} and \cite{Sandro2}. More specifically, in these papers, two field theory models of
dark energy interacting with dark matter were constructed. In both the models, dark matter was associated with a massive Dirac field, interacting via Yukawa coupling with a tachyon scalar field in one model and with a quintessence field in the other. However, instead of choosing a particular form for the potential $V\left(\varphi\right) $ of the scalar fields, this was implicitly fixed by imposing that the dark energy density
must match the energy density of the holographic dark energy. In this model, the dark energy density is given by $\rho_{de}=\frac{3M_{Pl}^{2}c^{2}}{L^{2}}$, where $M_{Pl}=\frac{1}{\sqrt{8\pi G}}$ is the reduced Planck mass, $c$ is a free parameter and $L$ is an infrared cutoff. It was demonstrated in \cite{Li} and \cite{Li2} that if one choose $L$ as the event horizon of the universe, the model reproduces the present period of accelerated expansion. Holographic dark energy models have been extensively studied in the literature, for a review and a list of references, see \cite{reviewholo}. It was demonstrated in \cite{taqholo}, \cite{taqholo2} and \cite{escalarholo} that there
are correspondences between quintessence, tachyon and holographic dark energy, in the noninteracting cases. The scalar fields in \cite{Sandro1} and \cite{Sandro2} were interacting, and in that cases the combination with the holographic dark energy in fact resulted in two new models of interacting dark energy.

However, there is a consistency problem, concerning causality, which would be
pointed in the holographic dark energy model: this depends on the event horizon
of the universe, and this in turn only exists if the period of accelerated expansion
is forever. Another model of dark energy, was proposed in \cite{Nade} on which again
$\rho_{de}=\frac{3M_{Pl}^{2}n^{2}}{L^{2}}$, but $L$ being now the conformal time,  $\eta\left(  t\right)  \equiv\int_{0}%
^{t}\frac{dt^{\prime}}{a\left(  t^{\prime}\right)  }$ ($n$ is again a
free parameter of order unity). This model does not have the consistency problem
mentioned, and possesses another advantage: because the initial value of relative
density of dark energy is not a free parameter, this model has one less parameter
than the holographic dark energy, possessing, in the noninteracting case, the same
number of free parameters as the $\Lambda CDM$. In this paper, we will construct two field
theory models of interacting dark energy, one in which the dark energy is associated
with the quintessence, and another in which the dark energy is the tachyon.
However, instead of choosing the potentials $V\left(  \varphi\right)  $, we will specify these implicitly,
by imposing that the energy density of the scalar fields,$\rho_{\varphi}$, must match the new
agegraphic dark energy density, $\rho
_{de}=\frac{3M_{Pl}^{2}n^{2}}{\eta^{2}}$. This was the same reasoning used
in \cite{Sandro1} and \cite{Sandro2} to construct the two models analyzed in that papers, but there
holographic dark energy was in place of the new agegraphic dark energy. Therefore,
now the models possess different dynamical properties, as the new agegraphic dark
energy model behaves itself different from the holographic dark energy, as already discussed in \cite{Nade}.
 Moreover, the models have no causality problem, and possess
one less parameter than before.
In this paper, we use the Natural Units system, in which $\hbar=c=k_{B}=1$.

\section{Interacting New Agegraphic Dark Energy}

It was argued in \cite{Karolyhazy} - \cite{Karolyhazy3} that
a distance $t$ in Minkowski space cannot be measured with accuracy better than%
\begin{equation}
\delta t=\lambda t_{p}^{2/3}t^{1/3}\text{ ,} \label{delta_t}%
\end{equation}
\ where $\lambda$ is a dimensionless constant of order unity, $t_{p}$ is the
reduced Planck time, given by $t_{p}=\frac{1}{M_{Pl}}$, being $M_{Pl}$ the
reduced Planck mass. Because the time-energy uncertainty relation, this
uncertainty on lenght measures implies that a region of size $\delta t^{3}$
possesses an energy content \cite{Maziashvili1} \cite{Maziashvili2}%
\begin{equation}
E_{\delta t^{3}}\sim t^{-1}\text{.} \label{energy_1}%
\end{equation}
Therefore, there is an energy density associated with the quantum fluctuations
of the space-time, given by%
\begin{equation}
\rho_{q}\sim\frac{E_{\delta t^{3}}}{\delta t^{3}}\sim\frac{1}{t_{p}^{2}t^{2}%
}\sim\frac{M_{Pl}^{2}}{t^{2}}\text{ .} \label{ro_q}%
\end{equation}
This energy density was associated with dark energy in \cite{Cai}. More
precisely, the dark energy density would be given by%
\begin{equation}
\rho_{DE}=\frac{3n^{2}M_{Pl}^{2}}{T^{2}}\text{ ,} \label{age_de}%
\end{equation}
where the time $t$ was identified with the age of the universe $T$ and $n$
is a dimensionless parameter of order unity. The resultant model of dark
energy was denominated \textit{agegraphic dark energy}. However, this model
has a subtlety \cite{Cai}, and in \cite{Nade} it was proposed that the age of
the universe $T$ be replaced by the conformal time $\eta$, that is,
\begin{equation}
\rho_{DE}=\frac{3n^{2}M_{Pl}^{2}}{\eta^{2}}\text{ ,} \label{nade_de}%
\end{equation}
where%
\begin{equation}
\eta\left(  t\right)  \equiv\int_{0}^{t}\frac{dt^{\prime}}{a\left(  t^{\prime
}\right)  } \label{conformal_time}%
\end{equation}
is the conformal time. The model of dark energy resulting was denominated
\textit{new agegraphic dark energy model}.

It is interesting to note that, from a different argumentation it was
obtained in \cite{Li} and \cite{Li2} the holographic dark energy model, whose
expression for the dark energy density is very similar to
(\ref{nade_de}), namely%
\begin{equation}
\rho_{DE}=\frac{3c^{2}M_{Pl}^{2}}{R_{h}^{2}}\text{ ,} \label{holo_de}%
\end{equation}
where $R_{h}$ is the event horizon, given by%
\begin{equation}
R_{h}=a\left(  t\right)  \int_{t}^{\infty}\frac{dt^{\prime}}{a(t^{\prime}%
)}\text{ .} \label{event_horizon2}%
\end{equation}

The reasoning used to construct both the models, although different, has in
common that quantum effects of gravity are incorporated for very small scales. Therefore, although we don't have a quantum gravity theory yet, the similarity of the expressions (\ref{nade_de})
and (\ref{holo_de}) perhaps suggests that we correctly incorporated some
universal property of quantum gravity.

For a universe composed by dark energy and dark matter in interaction, and
baryonic matter and radiation, the conservation equations are%

\begin{equation}
\dot{\rho}_{DE}+3H\rho_{DE}(\omega_{DE}+1)=Q\text{ ,} \label{conser_de}%
\end{equation}%
\begin{equation}
\dot{\rho}_{DM}+3H\rho_{DM}=-Q\text{ ,} \label{conser_dm}%
\end{equation}%
\begin{equation}
\dot{\rho}_{b}+3H\rho_{b}=0 \label{conserbaryon}%
\end{equation}
and%
\begin{equation}
\dot{\rho}_{r}+4H\rho_{r}=0\text{ ,} \label{conserrad}%
\end{equation}
where the dot represents derivative with respect to time, and $Q$ is the
interaction term. The Friedmann equation for a flat universe reads%

\begin{equation}
H^{2}=\frac{1}{3M_{Pl}^{2}}\left[  \rho_{DE}+\rho_{DM}+\rho_{b}+\rho
_{r}\right]  \text{ .} \label{friedmann_sc}%
\end{equation}

Using Eqs. (\ref{conser_de})-(\ref{friedmann_sc}), it is possible to rewrite
(\ref{conser_de}) as%
\begin{equation}
\overset{\centerdot}{\Omega}_{DE}=3H\Omega_{DE}\left[  -\left(  1-\Omega
_{DE}\right)  \omega_{DE}+\frac{\Omega_{r}}{3}\right]  +\frac{Q}{3M_{Pl}%
^{2}H^{2}} \label{evol_omde}%
\end{equation}

On the other hand, the energy density of the new agegraphic dark energy is
given by (\ref{nade_de}), which can be also written as%
\begin{equation}
\Omega_{DE}=\frac{n^{2}}{H^{2}\eta^{2}} \label{agegraphic_omega}%
\end{equation}

Deriving (\ref{nade_de}) with respect to time, and using
(\ref{conformal_time}) and (\ref{agegraphic_omega}), we have%
\begin{equation}
\dot{\rho}_{DE}=-H\rho_{DE}\frac{2\sqrt{\Omega_{DE}}}{na} \label{rodot}%
\end{equation}

Inserting (\ref{rodot}) in (\ref{conser_de}) we obtain%
\begin{equation}
\omega_{DE}=-1+\frac{2\sqrt{\Omega_{DE}}}{3na}+\frac{Q}{3H\rho_{DE}}
\label{eos_int}%
\end{equation}

The interaction term $Q$ is especified by the interacting dark energy model which is under consideration. In this work, we will construct two field theory models
of interacting dark energy.

\section{The models}

We consider the general action%
\begin{equation}
S=\int d^{4}x\sqrt{-g}\left\{  -\frac{M_{Pl}^{2}}{2}R+\mathcal{L}_{\varphi
}\left(  x\right)  +\frac{i}{2}[\bar{\Psi}\gamma^{\mu}\nabla_{\mu}\Psi
-\bar{\Psi}\overleftarrow{\nabla}_{\mu}\gamma^{\mu}\Psi]-(M-\beta\varphi
)\bar{\Psi}\Psi+\sum\limits_{j}\mathcal{L}_{j}\left(  x\right)  \right\}
\label{action}%
\end{equation}
where $M_{Pl}\equiv\left(  8\pi G\right)  ^{-1/2}$ is the reduced Planck mass,
$R$ is the curvature scalar, $\mathcal{L}_{\varphi}\left(  x\right)  $ is,
except for the coupling term, the Lagrangian density for the scalar field,
which we will identify with dark energy, $\Psi$ is a massive fermionic field,
which we will identify with dark matter, $\beta$ is the dimensionless coupling
constant and $\sum\limits_{j}\mathcal{L}_{j}\left(  x\right)  $ contains the
Lagrangian densities for the ramaining fields. Note that, in this work, we
will only consider an interaction of dark energy with dark matter. If there
was a coupling between the scalar field and baryonic matter, the corresponding
coupling constant $\beta_{b}$ should satisfy the solar system constraint
\cite{solarsystem}%
\begin{equation}
\beta_{b}\lesssim10^{-2}\text{ .} \label{ss}%
\end{equation}
We assume $\beta_{b}\equiv0$, which trivially satisfy the constraint (\ref{ss}).

We consider two kinds of scalar fields: the canonical scalar field, or
quintessence field, for which%
\begin{equation}
\mathcal{L}_{\varphi}\left(  x\right)  =\frac{1}{2}\partial_{\mu}%
\varphi\partial^{\mu}\varphi-V(\varphi)\text{ ,} \label{canonical}%
\end{equation}
and the tachyon scalar field, for which%
\begin{equation}
\mathcal{L}_{\varphi}\left(  x\right)  =-V(\varphi)\sqrt{1-\alpha\partial
^{\mu}\varphi\partial_{\mu}\varphi}\text{ ,} \label{tachyonic}%
\end{equation}
where $\alpha$ is a constant with dimension $MeV^{-4}$. Note that in both
cases, we assume a Yukawa coupling with the dark matter field $\Psi$.

\subsection{Quintessence field}

For the quintessence field, $\mathcal{L}_{\varphi}\left(  x\right)  $ in the
action (\ref{action}) is given by (\ref{canonical}). From a variational
principle, we obtain%

\begin{equation}
i\gamma^{\mu}\nabla_{\mu}\Psi-M^{\ast}\Psi=0\text{ ,} \label{dirac}%
\end{equation}%
\begin{equation}
i(\nabla_{\mu}\bar{\Psi})\gamma^{\mu}+M^{\ast}\bar{\Psi}=0\text{ ,}
\label{diracadj}%
\end{equation}
where $M^{\ast}\equiv M-\beta\varphi$, and%
\begin{equation}
\nabla_{\mu}\partial^{\mu}\varphi+\frac{dV(\varphi)}{d\varphi}=\beta\bar{\Psi
}\Psi\text{ .} \label{eqmov_scalar}%
\end{equation}
Eqs. (\ref{dirac}) and (\ref{diracadj}) are, respectively, the covariant Dirac
equation and its adjoint, in the case of a nonvanishing interaction between
the Dirac field and the scalar field $\varphi$. For homogeneous fields and
adopting the flat Friedmann-Robertson-Walker (FRW) metric, $g_{\mu\nu}%
$=diag$\left(  1,-a^{2}\left(  t\right)  ,-a^{2}\left(  t\right)
,-a^{2}\left(  t\right)  \right)  $, where $a^{2}\left(  t\right)  $ is the
scale factor, Eqs. (\ref{dirac}) and (\ref{diracadj}) lead to%
\[
\frac{d(a^{3}\bar{\Psi}\Psi)}{dt}=0
\]
which is equivalent to%
\begin{equation}
\bar{\Psi}\Psi=\bar{\Psi}_{i}\Psi_{i}\left(  \frac{a_{i}}{a}\right)  ^{3}
\label{conser_psibarpsi}%
\end{equation}
where the subscript ``$i$" denotes some initial time, and (\ref{eqmov_scalar})
reduces to%
\begin{equation}
\ddot{\varphi}+3H\dot{\varphi}+\frac{dV(\varphi)}{d\varphi}=\beta\bar{\Psi
}\Psi\text{ ,} \label{homoscalar}%
\end{equation}
where $H\equiv\frac{\dot{a}}{a}$ is the Hubble parameter.

From the energy-momentum tensor, we get%
\begin{align}
\rho_{\varphi}  &  =\frac{1}{2}\dot{\varphi}^{2}+V(\varphi)\text{
,}\label{rofiscalar}\\
P_{\varphi}  &  =\frac{1}{2}\dot{\varphi}^{2}-V(\varphi)\text{ ,}%
\label{pfiscalar}\\
\rho_{\Psi}  &  =M^{\ast}\bar{\Psi}\Psi\text{ ,}\label{ropsi}\\
P_{\Psi}  &  =0\text{ .}\nonumber
\end{align}
From (\ref{rofiscalar}) and (\ref{pfiscalar}) we have $\omega_{\varphi}%
\equiv\frac{P_{\varphi}}{\rho_{\varphi}}=\frac{\frac{1}{2}\dot{\varphi}%
^{2}-V(\varphi)}{\frac{1}{2}\dot{\varphi}^{2}+V(\varphi)}$. Differentianting
(\ref{rofiscalar}) and (\ref{ropsi}) with respect to time and using
(\ref{conser_psibarpsi}) and (\ref{homoscalar}), we obtain%

\begin{equation}
\dot{\rho}_{\varphi}+3H\rho_{\varphi}(\omega_{\varphi}+1)=\beta\overset
{\centerdot}{\varphi}\bar{\Psi}_{i}\Psi_{i}\left(  \frac{a_{i}}{a}\right)
^{3} \label{conser_de_fi}%
\end{equation}
and%
\begin{equation}
\dot{\rho}_{\Psi}+3H\rho_{\Psi}=-\beta\overset{\centerdot}{\varphi}\bar{\Psi
}_{i}\Psi_{i}\left(  \frac{a_{i}}{a}\right)  ^{3}\text{ .}
\label{conser_dm_psi}%
\end{equation}

Comparing (\ref{conser_de}) and (\ref{conser_dm}) with (\ref{conser_de_fi}) and (\ref{conser_dm_psi}), we see that%
\begin{equation}
Q=\beta\overset{\centerdot}{\varphi}\bar{\Psi}_{i}\Psi_{i}\left(  \frac{a_{i}%
}{a}\right)  ^{3} \label{Q_term}%
\end{equation}

Remembering that $\rho_{\Psi i}=3M_{Pl}^{2}H_{i}^{2}\Omega_{\Psi i}$ and using
(\ref{ropsi}), we have%
\begin{equation}
\bar{\Psi}_{i}\Psi_{i}=\frac{3M_{Pl}^{2}H_{i}^{2}\Omega_{\Psi i}}%
{M-\beta\varphi_{i}}\text{ ,} \label{psibarpsi_i}%
\end{equation}
where $\Omega_{\Psi i}$ is the initial relative energy density of the dark
matter, $H_{i}$ is the initial value of the Hubble parameter, and $\varphi
_{i}$ is the initial value of the quintessence field. From (\ref{rofiscalar})
and (\ref{pfiscalar}) we have
\begin{equation}
\overset{\centerdot}{\varphi}=sign[\overset{\centerdot}{\varphi}]\sqrt
{3}M_{Pl}H\sqrt{\Omega_{\varphi}\left(  1+\omega_{\varphi}\right)  }\text{ .}
\label{fidot}%
\end{equation}

Substituting (\ref{psibarpsi_i}) and (\ref{fidot}) in (\ref{Q_term}), we have%
\begin{equation}
Q=sign[\overset{\centerdot}{\varphi}]\delta M_{Pl}\frac{H_{i}^{2}}{H^{2}}%
\sqrt{3}\Omega_{\Psi i}\sqrt{\Omega_{\varphi}\left(  1+\omega_{\varphi
}\right)  }\left(  \frac{a_{i}}{a}\right)  ^{3}\text{ ,} \label{Q_term2}%
\end{equation}
where we have defined the effective coupling constant
\begin{equation}
\delta\equiv\frac{\beta}{M-\beta\varphi_{i}}\text{ .} \label{delta-quint}%
\end{equation}

Note that $sign[\overset{\centerdot}{\varphi}]$ is in fact arbitrary, as it
can be changed by redefinitions of the quintessence field, $\varphi
\rightarrow-\varphi$, and of the coupling constant $\beta\rightarrow-\beta$.
Substituting (\ref{Q_term2}) in (\ref{evol_omde}) we have%
\begin{equation}
\frac{d\Omega_{\varphi}}{dz}=\frac{3\Omega_{\varphi}}{1+z}\left\{  \left(
1-\Omega_{\varphi}\right)  \omega_{\varphi}-\frac{\Omega_{r}}{3}-\sqrt
{\frac{2}{3}}\gamma_{q}\sqrt{1+\omega_{\varphi}}\right\}  \text{ ,}
\label{evol_omfi_z}%
\end{equation}

where%
\begin{equation}
\gamma_{q}\left(  z\right)  \equiv\frac{\delta M_{Pl}}{\sqrt{2}}\left(
\frac{H_{i}}{H}\right)  ^{2}\frac{\Omega_{\Psi i}}{\sqrt{\Omega_{\varphi}}%
}\left(  \frac{1+z}{1+z_{i}}\right)  ^{3}\text{ .} \label{gama_quint}%
\end{equation}

Note that we rewrite the evolution equation for $\Omega_{\varphi}$ in terms
of redshift $z$.

Inserting (\ref{Q_term2}) in (\ref{eos_int}), we have%
\begin{equation}
\omega_{\varphi}=-1+\frac{2\sqrt{\Omega_{\varphi}}}{3n}\left(  1+z\right)
+\sqrt{\frac{2}{3}}\gamma_{q}\sqrt{1+\omega_{\varphi}}\text{ .}
\label{eos_int2}%
\end{equation}

Solving for $\omega_{\varphi}$, we obtain%
\begin{equation}
\omega_{\varphi}\left(  z\right)  =-1+\frac{2\sqrt{\Omega_{\varphi}}}%
{3n}\left(  1+z\right)  +\frac{\gamma_{q}}{3}\left[  \gamma_{q}+\sqrt
{\gamma_{q}^{2}+\frac{4\sqrt{\Omega_{\varphi}}}{n}\left(  1+z\right)
}\right]  \text{ .} \label{eos_int_quint}%
\end{equation}

In an entirely analogue manner done to (\ref{conser_de}), one can rewrite
(\ref{conser_dm})-(\ref{conserrad}) as
\begin{equation}
\frac{d\Omega_{\Psi}}{dz}=-\frac{3}{1+z}\left[  \Omega_{\Psi}\left(
\Omega_{\varphi}\omega_{\varphi}+\frac{\Omega_{r}}{3}\right)  -\sqrt{\frac
{2}{3}}\gamma_{q}\Omega_{\varphi}\sqrt{1+\omega_{\varphi}}\right]  \text{ ,}
\label{evol_ompsi_z}%
\end{equation}

\begin{equation}
\frac{d\Omega_{b}}{dz}=-\frac{3\Omega_{b}}{1+z}\left[  \Omega_{\varphi}%
\omega_{\varphi}+\frac{\Omega_{r}}{3}\right]  \text{ and} \label{evol_omb_z}%
\end{equation}

\begin{equation}
\frac{d\Omega_{r}}{dz}=-\frac{3\Omega_{r}}{1+z}\left[  \Omega_{\varphi}%
\omega_{\varphi}+\frac{\Omega_{r}}{3}-\frac{1}{3}\right]  \text{ .}
\label{evol_omr_z}%
\end{equation}

Evidently, from (\ref{evol_omfi_z}), (\ref{evol_ompsi_z})-(\ref{evol_omr_z}), only three are independent, as for a flat universe,
$\Omega_{\varphi}+\Omega_{\Psi}+\Omega_{b}+\Omega_{r}=1$.

From (\ref{conserrad}) we have%
\begin{equation}
\rho_{r}=\rho_{ri}\left(  \frac{1+z}{1+z_{i}}\right)  ^{4}\text{ .}
\label{ror_zi}%
\end{equation}
So%

\begin{equation}
\rho_{ri}=\rho_{r0}\left(  \frac{1+z_{i}}{1+z_{0}}\right)  ^{4}\text{ ,}
\label{ror_i}%
\end{equation}
where the subscript ``$0$" denotes the quantities today. $\rho_{r0}=\left(
1+0.2271N_{eff}\right)  \rho_{\gamma0}$, where $N_{eff}=3.04$ is the effective
number of relativistic degrees of freedom, and $\rho_{\gamma0}=\frac{\pi^{2}%
}{15}T_{CMB}^{4}$ is the energy density of photons, $T_{CMB}=2.725K$ is the
CMB temperature today. Remembering that $\rho_{ri}=3M_{Pl}^{2}H_{i}^{2}%
\Omega_{ri}$ and from (\ref{ror_i}), we have%
\begin{equation}
H_{i}=\lambda\frac{\left(  1+z_{i}\right)  ^{2}}{\sqrt{\Omega_{ri}}}\text{ ,}
\label{Hi}%
\end{equation}

where
\begin{equation}
\lambda\equiv\frac{\pi}{3M_{Pl}}\sqrt{\frac{1+0.2271N_{eff}}{5}}T_{CMB}%
^{2}\text{ .} \label{lambda}%
\end{equation}

On the other hand, from (\ref{ror_zi}) it is possible to write the Hubble
parameter as%
\begin{equation}
H\left(  z\right)  =H_{i}\sqrt{\frac{\Omega_{ri}}{\Omega_{r}}}\left(
\frac{1+z}{1+z_{i}}\right)  ^{2} \label{Hz1}%
\end{equation}

or, using (\ref{Hi}),%
\begin{equation}
H\left(  z\right)  =\lambda\frac{\left(  1+z\right)  ^{2}}{\sqrt{\Omega_{r}}%
}\text{ .} \label{Hz2}%
\end{equation}

According to \cite{Nade}, the initial value $\Omega_{\varphi i}$ is not a free
parameter, but, in the radiation era, it is related with the parameter $n$ as%
\begin{equation}
\Omega_{\varphi i}=\frac{n^{2}}{(1+z_{i})^{2}}\text{ ,} \label{omfi_i}%
\end{equation}
where $z_{i}$ is some redshift for which the universe was in the radiation era.

So, the evolution with redshift $z$ of all quantities of the model are
determined by three of the Eqs. (\ref{evol_omfi_z}), (\ref{evol_ompsi_z}-(\ref{evol_omr_z}), with $\omega_{\varphi}$ given by
(\ref{eos_int_quint}), $\gamma_{q}\left(  z\right)  $ and $H\left(  z\right)
$ given by (\ref{gama_quint}) and (\ref{Hz2}) respectively. The free
parameters of the model are $\delta$, $n$, $\Omega_{\Psi i}$ and $\Omega_{bi}%
$. (The initial condition $\varphi_{i}$ is in fact arbitrary, as only the
effective coupling constant $\delta$ is constrained by the observational
data.) It is interesting to note that for the noninteracting case,
$\delta=0$, the model has three free parameters, $n$, $\Omega_{\Psi i}$ and
$\Omega_{bi}$, the same number of free parameters as the $\Lambda CDM$, for
which the free parameters are $\Omega_{\Lambda i}$, $\Omega_{\Psi i}$ and
$\Omega_{bi}$.

The relation (\ref{fidot}) can be rewritten in terms of redshift as%
\begin{equation}
\frac{d\varphi}{dz}=-\frac{\sqrt{3}M_{Pl}\sqrt{\Omega_{\varphi}\left(
z\right)  \left(  1+\omega_{\varphi}\left(  z\right)  \right)  }}{1+z}\text{
.} \label{dfi_dz}%
\end{equation}

From (\ref{rofiscalar}) and (\ref{fidot}) we have%
\begin{equation}
V\left(  z\right)  =3M_{Pl}^{2}H^{2}\frac{\Omega_{\varphi}\left(  z\right)
\left(  1-\omega_{\varphi}\left(  z\right)  \right)  }{2}\text{ .} \label{Vz}%
\end{equation}

From (\ref{Vz}) and (\ref{dfi_dz}) it is possible to compute $V\left(
\varphi\right)  $. Hereafter, we denote the \textit{Interacting Quintessence
New Agegraphic Dark Energy Model }simply as IQNADE.

\subsection{Tachyon field}

In the case of dark energy modeled as the tachyon scalar field, $\mathcal{L}%
_{\varphi}\left(  x\right)  $ in the action (\ref{action}) is given by
(\ref{tachyonic}). From a variational principle, we obtain%
\begin{equation}
i\gamma^{\mu}\nabla_{\mu}\Psi-M^{\ast}\Psi=0\text{ ,} \label{dirac_tac}%
\end{equation}%
\begin{equation}
i(\nabla_{\mu}\bar{\Psi})\gamma^{\mu}+M^{\ast}\bar{\Psi}=0\text{ ,}
\label{diracadj_tac}%
\end{equation}
where $M^{\ast}\equiv M-\beta\varphi$, and%
\begin{equation}
\nabla_{\mu}\partial^{\mu}\varphi+\alpha\frac{\partial^{\mu}\varphi
(\nabla_{\mu}\partial_{\sigma}\varphi)\partial^{\sigma}\varphi}{1-\alpha
\partial_{\mu}\varphi\partial^{\mu}\varphi}+\frac{1}{\alpha}\frac
{dlnV(\varphi)}{d\varphi}=\frac{\beta\bar{\Psi}\Psi}{\alpha V(\varphi)}%
\sqrt{1-\alpha\partial^{\mu}\varphi\partial_{\mu}\varphi}\text{ .}
\label{eqmov_taquions}%
\end{equation}
Equations (\ref{dirac_tac}) and (\ref{diracadj_tac}) are the interacting covariant
Dirac equation and its adjoint, respectively, i. e., (\ref{dirac_tac}) and
(\ref{diracadj_tac}) are almost the same as eqs. (\ref{dirac}) and
(\ref{diracadj}), the only difference is that the scalar field $\varphi$ in
$M^{\ast}$ now is the tachyon field. For homogeneous fields and adopting the
flat FRW metric, (\ref{eqmov_taquions}) reduces to%
\begin{equation}
\ddot{\varphi}=-(1-\alpha\dot{\varphi}^{2})\left[  \frac{1}{\alpha}%
\frac{dlnV(\varphi)}{d\varphi}+3H\dot{\varphi}-\frac{\beta\bar{\Psi}\Psi
}{\alpha V(\varphi)}\sqrt{1-\alpha\dot{\varphi}^{2}}\right]  \text{ ,}
\label{homotaq}%
\end{equation}
whereas for the fermions, the equations of motion will reduce to eq.
(\ref{conser_psibarpsi}), as already obtained above:%
\begin{equation}
\bar{\Psi}\Psi=\bar{\Psi}_{i}\Psi_{i}\left(  \frac{a_{i}}{a}\right)
^{3}\text{ .} \label{psibarpsitaq}%
\end{equation}

From the energy-momentum tensor, we get%
\begin{align}
\rho_{\varphi}  &  =\frac{V(\varphi)}{\sqrt{1-\alpha\dot{\varphi}^{2}}}\text{
,}\label{rofi}\\
P_{\varphi}  &  =-V(\varphi)\sqrt{1-\alpha\dot{\varphi}^{2}}\text{
,}\label{pfi}\\
\rho_{\Psi}  &  =M^{\ast}\bar{\Psi}\Psi\text{ ,}\label{ropsitaq}\\
P_{\Psi}  &  =0\text{ .}\nonumber
\end{align}
From (\ref{rofi}) and (\ref{pfi}) we have
\begin{equation}
\omega_{\varphi}\equiv\frac{P_{\varphi}}{\rho_{\varphi}}=\alpha\dot{\varphi
}^{2}-1\text{ .} \label{wfi}%
\end{equation}
Differentiating (\ref{rofi}) and (\ref{ropsitaq}) with respect to time and using
(\ref{homotaq}) and (\ref{psibarpsitaq}), we get%

\begin{equation}
\dot{\rho}_{\varphi}+3H\rho_{\varphi}(\omega_{\varphi}+1)=\beta\overset
{\centerdot}{\varphi}\bar{\Psi}_{i}\Psi_{i}\left(  \frac{a_{i}}{a}\right)
^{3} \label{conser_rotac}%
\end{equation}
and%
\begin{equation}
\dot{\rho}_{\Psi}+3H\rho_{\Psi}=-\beta\overset{\centerdot}{\varphi}\bar{\Psi
}_{i}\Psi_{i}\left(  \frac{a_{i}}{a}\right)  ^{3}\text{ ,}
\label{conser_psitac}%
\end{equation}
where the dot represents derivative with respect to time.

Note that the interaction term is of the same form as before,
\begin{equation}
Q=\beta\overset{\centerdot}{\varphi}\bar{\Psi}_{i}\Psi_{i}\left(  \frac{a_{i}%
}{a}\right)  ^{3}\text{ .} \label{Q_term3}%
\end{equation}
However, the scalar field now is the tachyon, its behaviour been determined by
(\ref{homotaq}). Defining $\phi\equiv\sqrt{\alpha}\varphi$, from (\ref{wfi}),
we have%
\begin{equation}
\overset{\centerdot}{\phi}=sign[\overset{\centerdot}{\phi}]\sqrt
{1+\omega_{\phi}}\text{ .} \label{fidot2}%
\end{equation}
As before, we have%
\begin{equation}
\bar{\Psi}_{i}\Psi_{i}=\frac{3M_{Pl}^{2}H_{i}^{2}\Omega_{\Psi i}}%
{M-\frac{\beta}{\sqrt{\alpha}}\phi_{i}}\text{ .} \label{psibarpsitaqi}%
\end{equation}
Substituting (\ref{fidot2}) and (\ref{psibarpsitaqi}) in (\ref{Q_term3}), we
have%
\begin{equation}
Q=sign[\overset{\centerdot}{\phi}]\delta3M_{Pl}^{2}H_{i}^{2}\Omega_{\Psi
i}\sqrt{1+\omega_{\phi}}\left(  \frac{a_{i}}{a}\right)  ^{3}\text{ ,}
\label{Qterm_taq}%
\end{equation}
where%
\begin{equation}
\delta\equiv\frac{\frac{\beta}{M\sqrt{\alpha}}}{1-\frac{\beta}{M\sqrt{\alpha}%
}\phi_{i}}\text{ .} \label{delta_taq}%
\end{equation}
As before, $sign[\overset{\centerdot}{\phi}]$ is in fact arbitrary, as it can
be changed by redefinitions of the tachyon field, $\phi\rightarrow-\phi$, and
of the coupling constant $\beta\rightarrow-\beta$. Substituting
(\ref{Qterm_taq}) in (\ref{evol_omde}) we have%
\begin{equation}
\frac{d\Omega_{\phi}}{dz}=\frac{3\Omega_{\phi}}{1+z}\left\{  \left(
1-\Omega_{\phi}\right)  \omega_{\phi}-\frac{\Omega_{r}}{3}-\sqrt{\frac{2}{3}%
}\gamma_{t}\sqrt{1+\omega_{\phi}}\right\}  \text{ ,}%
\end{equation}
where%
\begin{equation}
\gamma_{t}\left(  z\right)  =\frac{1}{\sqrt{6}}\delta\frac{H_{i}^{2}}{H^{3}%
}\frac{\Omega_{\Psi i}}{\Omega_{\phi}}\left(  \frac{1+z}{1+z_{i}}\right)
^{3}\text{ .} \label{gama_taq}%
\end{equation}
In an analogue manner as done for quintessence, we obtain%
\begin{equation}
\omega_{\phi}\left(  z\right)  =-1+\frac{2\sqrt{\Omega_{\phi}}}{3n}\left(
1+z\right)  +\frac{\gamma_{t}}{3}\left[  \gamma_{t}+\sqrt{\gamma_{t}^{2}%
+\frac{4\sqrt{\Omega_{\phi}}}{n}\left(  1+z\right)  }\right]  \text{ .}
\label{wfi_taq}%
\end{equation}
As before, the Friedmann equation reads%
\begin{equation}
H\left(  z\right)  =\lambda\frac{\left(  1+z\right)  ^{2}}{\sqrt{\Omega_{r}}%
}\text{ ,}%
\end{equation}
where%
\begin{equation}
\lambda\equiv\frac{\pi}{3M_{Pl}}\sqrt{\frac{1+0.2271N_{eff}}{5}}T_{CMB}%
^{2}\text{ .}%
\end{equation}
Again, the initial value $\Omega_{\phi i}$ is not a free parameter, but is
determined by $n$ as%
\begin{equation}
\Omega_{\phi i}=\frac{n^{2}}{(1+z_{i})^{2}}\text{ .}%
\end{equation}
So the interacting tachyonic agegraphic dark energy model possesses four free
parameters: $\delta$, $n$, $\Omega_{\Psi i}$ and $\Omega_{bi}$, which must be
determined from comparison of the model with observational data. Again,
$\phi_{i}$ is arbitray, as only the effective coupling constant $\delta$ is
constrained by the data.

We can obtain the evolution of $\phi$ with redshift as%
\begin{equation}
\frac{d\phi}{dz}=-\frac{\sqrt{1+\omega_{\phi}\left(  z\right)  }}{H\left(
z\right)  \left(  1+z\right)  }\text{ .} \label{dfi_dz_taq}%
\end{equation}
From (\ref{rofi}), the potential can be written as%
\begin{equation}
V\left(  z\right)  =3M_{Pl}^{2}H^{2}\Omega_{\phi}\left(  z\right)
\sqrt{-\omega_{\phi}\left(  z\right)  }\text{ .} \label{potential}%
\end{equation}
From (\ref{potential}) and (\ref{dfi_dz_taq}), it is possible to compute $V(\phi)$
for the tachyon field. Hereafter, we will refer to \textit{Interacting
Tachyonic New Agegraphic Dark Energy Model} as ITNADE.

It is interesting to note that both interacting models, IQNADE and ITNADE,
in the noninteracting case, $\delta=0$, will be reduced to the NADE model. In
other words, we will obtain reconstructions of NADE from the quintessence and
tachyon fields, as already obtained in \cite{quintreconstruction} - \cite{tachyonreconstruction}.

\section{Constraints from observational data}

We include four sets of observational data: the 1048 SNIa data from the
Pantheon sample \cite{Scolnic}, 9 baryonic acoustic oscillations (BAO) data as compiled, for instance, in
\cite{BAO} and \cite{BAO2}, measurements of the Hubble parameter in 31 different
redshifts, as compiled, for instance, in \cite{Hubble}, and the CMB distance
priors from Planck 2015 data \cite{PriorsPLANCK2015} \footnote{The distance priors obtained from Planck 2018 data has already been published \cite{PriorsPLANCK2018}. However, the constraints on the distance priors derived from Planck 2018 data are only slightly improved - at most $8\%$ - compared to those derived from Planck 2015 data \cite{PriorsPLANCK2018}, so that the use of distance priors from Planck 2015 data must not have significantly affected the inference of the parameters of the models studied here, mainly because they were used joint with other three data sets.}.

We compare our theoretical predictions for the distance modulus at redshift
$z$, $\mu(z)$, with the 1048 observational values of $\mu$ of the Pantheon
sample \cite{Scolnic}. The theoretical distance modulus is defined as
\begin{equation}
\mu(z)=5log_{10}\left[  c(1+z)\int_{0}^{z}\frac{dz^{\prime}}{H(z^{\prime}%
)}\right]  +15\text{ .} \label{muteo}%
\end{equation}

We compute the quantity
\begin{equation}
\chi_{SN}^{2}=%
{\displaystyle\sum\limits_{ij}}
\left(  \mu_{i}^{th}-\mu_{i}^{data}\right)  C_{ij}^{-1(Pantheon)}\left(  \mu_{j}%
^{th}-\mu_{j}^{data}\right)  \text{ ,} \label{chisn}%
\end{equation}
where $\mu^{th}$ are the predicted model values calculated using
(\ref{muteo}), and $\mu^{data}$ are the observational values of the Pantheon
sample. $C_{ij}^{-1(Pantheon)}$ is the inverse of the covariance matrix for the Pantheon sample.

The Planck distance priors summarize the information of temperature power
spectrum of CMB. These includes the ``shift parameter" $R$, the ``acoustic
scale" $l_{A}$ and the physical energy density of baryonic matter today,
$\Omega_{b0}h^{2}$. These quantities are very weakly model-dependent
\cite{liR} \cite{liR2}. $R$ and $l_{A}$ are given by%
\[
R=\sqrt{\Omega_{m0}}H_{0}r\left(  z_{\ast}\right)
\]
and%
\[
l_{A}=\pi\frac{r\left(  z_{\ast}\right)  }{r_{s}\left(  z_{\ast}\right)
}\text{ ,}%
\]
where $r\left(  z_{\ast}\right)  $ is the comoving distance to redshift of
last scattering $z_{\ast}$, $\ r_{s}\left(  z_{\ast}\right)  $ is the comoving
sound horizon at $z_{\ast}$, $\Omega_{m0}=\Omega_{DM0}+\Omega_{b0}$, the total
energy density of matter today (dark matter plus baryonic matter) and $H_{0}$
is the Hubble parameter today. For a flat universe, $r\left(z\right) $ and $r_{s}\left(z\right) $ are given by%
\begin{equation}
r\left(  z\right)  =\int_{0}^{z}\frac{dz}{H(z)} \label{distancia_comovel}%
\end{equation}
and%
\begin{equation}
r_{s}\left(  z\right)  =\int_{z}^{\infty}\frac{dz}{H(z)\sqrt{3\left(
1+\bar{R}_{b}/(1+z)\right)  }}\text{ ,} \label{sound_horizon}%
\end{equation}
where $\bar{R}_{b}/(1+z)=3\Omega_{b}/\left(  4\Omega_{\gamma}\right)  $. For
the redshift of decoupling $z_{\ast}$ we use the fitting function proposed by
Hu and Sugiyama \cite{sugiyama}:%
\[
z_{\ast}=1048\left[  1+0.00124\left(  \Omega_{b0}h^{2}\right)  ^{-0.738}%
\right]  \left[  1+g_{1}\left(  \Omega_{m0}h^{2}\right)  ^{g_{2}}\right]
\text{ ,}%
\]
where%
\[
g_{1}=\frac{0.0783\left(  \Omega_{b0}h^{2}\right)  ^{-0.238}}{1+39.5\left(
\Omega_{b0}h^{2}\right)  ^{0.763}}%
\]
and%
\[
g_{2}=\frac{0.560}{1+21.1\left(  \Omega_{b0}h^{2}\right)  ^{1.81}}\text{ .}%
\]

Table \textbf{1} shows the Planck distance information \cite{PriorsPLANCK2015} used in this work.

\begin{center}
\textbf{Table} \textbf{1}: Planck distance information from Planck 2015 data.

\bigskip%

\begin{tabular}
[c]{|l|l|}\hline
$R$ & $1.7448$\\\hline
$l_{A}$ & $301.460$\\\hline
$\Omega_{b0}h^{2}$ & $0.02240$\\\hline
\end{tabular}

\end{center}

The inverse of the covariance matrix associated with these data is given below
\cite{PriorsPLANCK2015}%

\[
C_{ij}^{-1\left(  PLANCK\right)  }=\left(
\begin{tabular}
[c]{lll}%
$84362.33$ & $-1314.56$ & $1650925.67$\\
$-1314.56$ & $157.90$ & $6186.87$\\
$1650925.67$ & $6186.87$ & $74320938.55$%
\end{tabular}
\ \right)
\]

Thus we add to $\chi_{tot}^{2}$ the term%
\[
\chi_{CMB}^{2}=%
{\displaystyle\sum\limits_{ij}}
\left(  x_{i}^{th}-x_{i}^{data}\right)  C_{ij}^{-1\left(  PLANCK\right)
}\left(  x_{j}^{th}-x_{j}^{data}\right)  \text{ ,}%
\]
where $x=\left(  l_{A},R,\Omega_{b0}h^{2}\right)  $ is the parameter vector.

BAO are described in terms of the cosmological distances%

\begin{equation}
D_{V}\left(  z\right)  =c\left[  \frac{z}{H\left(  z\right)  }\left(  \int
_{0}^{z}\frac{dz^{\prime}}{H\left(  z^{\prime}\right)  }\right)  ^{2}\right]
^{1/3}\text{ ,} \label{Dv}%
\end{equation}%
\begin{equation}
D_{A}=\frac{c}{1+z}\int_{0}^{z}\frac{dz^{\prime}}{H\left(  z^{\prime}\right)
}\text{ ,} \label{DA}%
\end{equation}%
\begin{equation}
D_{H}=\frac{c}{H}\text{ .} \label{DH}%
\end{equation}

The observational values of BAO which we use in this work are given in terms
of the quotients of (\ref{Dv}) - (\ref{DH}) with $r_{d}$, the
comoving sound horizon at $z_{d}$, the redshift of the drag epoch. The
theoretical value of $r_{d}$ is calculated using (\ref{sound_horizon}) with
$z=z_{d}$, where $z_{d}$ is calculated using the fitting function proposed by
Eisenstein and Hu \cite{Eisenstein}:%
\[
z_{d}=1291\frac{\left(  \Omega_{m0}h^{2}\right)  ^{0.251}}{1+0.659\left(
\Omega_{m0}h^{2}\right)  ^{0.828}}\left[  1+b_{1}\left(  \Omega_{b0}%
h^{2}\right)  ^{b_{2}}\right]  \text{ ,}%
\]

\[
b_{1}=0.313\left(  \Omega_{m0}h^{2}\right)  ^{-0.419}\left[  1+0.607\left(
\Omega_{m0}h^{2}\right)  ^{0.674}\right]  \text{ ,}%
\]

\[
b_{2}=0.238\left(  \Omega_{m0}h^{2}\right)  ^{0.223}\text{ .}%
\]

The observational values of BAO used here are given in tables \textbf{2} and
\textbf{3 }below, and were compiled in \cite{BAO2}.

\begin{center}
\textbf{Table} \textbf{2}: Isotropic BAO scale measurements.

\bigskip%

\begin{tabular}
[c]{|c|c|c|}\hline
$z$ & $d_{i}^{iso}$ & \\\hline
$0.106$ & $\frac{D_{V}\left(  0.106\right)  }{r_{d}}=2.98\pm0.13$ &
\cite{Beutler}\\\hline
$0.15$ & $\frac{D_{V}\left(  0.15\right)  }{r_{d}}=4.47\pm0.17$ &
\cite{Ross}\\\hline
$1.52$ & $\frac{D_{V}\left(  1.52\right)  }{r_{d}}=26.1\pm1.1$ &
\cite{Ata}\\\hline
\end{tabular}

\bigskip

\bigskip

\textbf{Table} \textbf{3}: Anisotropic BAO scale measurements.

\bigskip%

\begin{tabular}
[c]{|c|c|c|}\hline
z & d$_{i}^{aniso}$ & \\\hline
0.38 & $\frac{D_{A}\left(  0.38\right)  }{r_{d}}=7.42$ & \cite{Alam}\\\hline
0.38 & $\frac{D_{H}\left(  0.38\right)  }{r_{d}}=24.97$ & \cite{Alam}\\\hline
0.51 & $\frac{D_{A}\left(  0.51\right)  }{r_{d}}=8.85$ & \cite{Alam}\\\hline
0.51 & $\frac{D_{H}\left(  0.51\right)  }{r_{d}}=22.31$ & \cite{Alam}\\\hline
0.61 & $\frac{D_{A}\left(  0.61\right)  }{r_{d}}=9.69$ & \cite{Alam}\\\hline
0.61 & $\frac{D_{H}\left(  0.61\right)  }{r_{d}}=20.49$ & \cite{Alam}\\\hline
\end{tabular}

\end{center}

The $\chi_{BAO}^{2}$ is given by%

\begin{equation}
\chi_{BAO}^{2}=\chi_{iso}^{2}+\chi_{aniso}^{2} \label{chiqBAO}%
\end{equation}
with%
\begin{equation}
\chi_{iso}^{2}=\sum_{i}\left(  \frac{d_{i}^{iso}-d_{i(th)}^{iso}}{\sigma_{i}%
}\right)  ^{2} \label{chiqBAOiso}%
\end{equation}
and%

\begin{equation}
\chi_{aniso}^{2}=\sum_{ij}\left(  d_{i}^{aniso}-d_{i(th)}^{aniso}\right)
C_{ij}^{-1\left(  BAO\right)  }\left(  d_{j}^{aniso}-d_{j(th)}^{aniso}\right)
\text{ ,} \label{chiqBAOaniso}%
\end{equation}
where $C_{ij}^{-1\left(  BAO\right)  }$ is the inverse of the covariance
matrix for the anisotropic BAO \cite{BAO}, given by%

\[
C_{ij}^{-1\left(  BAO\right)  }=\left(
\begin{tabular}
[c]{llllll}%
$100.412$ & $7.19968$ & $-44.2237$ & $-5.43336$ & $4.73801$ & $1.09265$\\
$7.19968$ & $2.82564$ & $-3.61277$ & $-1.77055$ & $0.707386$ & $0.32102$\\
$-44.2237$ & $-3.61277$ & $106.03$ & $11.7756$ & $-38.2942$ & $-5.97272$\\
$-5.43336$ & $-1.77055$ & $11.7756$ & $6.1121$ & $-4.76135$ & $-3.01821$\\
$4.73801$ & $0.707386$ & $-38.2942$ & $-4.76135$ & $66.2442$ & $9.29217$\\
$1.09265$ & $0.32102$ & $-5.97272$ & $-3.01821$ & $9.29217$ & $6.22445$%
\end{tabular}
\ \ \right)
\]

We also inlcude values for the Hubble parameter $H$ in 31 redshifts. These
data are compiled, e. g., in \cite{Hubble}, and are listed in table \textbf{4}.

\begin{center}
\textbf{Table 4}: The $H\left(  z\right)  $ data. The values of $H$ are in
$\frac{km}{sMpc}$.

\bigskip%

\begin{tabular}
[c]{|l|l|l|l|l|l|}\hline
$z$ & $H(1\sigma)$ & Ref. & $z$ & $H(1\sigma)$ & Ref.\\\hline
$0.07$ & $69.0(19.6)$ & \cite{Zhang2014} & $0.4783$ & $80.9(9.0)$ &
\cite{Moresco2016}\\\hline
$0.09$ & $69.0(12.0)$ & \cite{Simon2005} & $0.48$ & $97.0(62.0)$ &
\cite{Stern2010}\\\hline
$0.12$ & $69.0(12.0)$ & \cite{Zhang2014} & $0.5929$ & $104.0(13.0)$ &
\cite{Moresco2012}\\\hline
$0.17$ & $83.0(8.0)$ & \cite{Simon2005} & $0.6797$ & $92.0(8.0)$ &
\cite{Moresco2012}\\\hline
$0.1791$ & $75.0(4.0)$ & \cite{Moresco2012} & $0.7812$ & $105.0(12.0)$ &
\cite{Moresco2012}\\\hline
$0.1993$ & $75.0(5.0)$ & \cite{Moresco2012} & $0.8754$ & $125.0(17.0)$ &
\cite{Moresco2012}\\\hline
$0.2$ & $72.9(29.6)$ & \cite{Zhang2014} & $0.88$ & $90.0(40.0)$ &
\cite{Stern2010}\\\hline
$0.27$ & $77.0(14.0)$ & \cite{Simon2005} & $0.9$ & $117.0(23.0)$ &
\cite{Simon2005}\\\hline
$0.28$ & $88.8(36.6)$ & \cite{Zhang2014} & $1.037$ & $154.0(20.0)$ &
\cite{Moresco2012}\\\hline
$0.3519$ & $83.0(14.0)$ & \cite{Moresco2012} & $1.3$ & $168.0(17.0)$ &
\cite{Simon2005}\\\hline
$0.3802$ & $83.0(13.5)$ & \cite{Moresco2016} & $1.363$ & $160.0(33.0)$ &
\cite{Moresco2015}\\\hline
$0.4$ & $95.0(17.0)$ & \cite{Simon2005} & $1.43$ & $177.0(18.0)$ &
\cite{Simon2005}\\\hline
$0.4004$ & $77.0(10.2)$ & \cite{Moresco2016} & $1.53$ & $140.0(14.0)$ &
\cite{Simon2005}\\\hline
$0.4247$ & $87.1(11.2)$ & \cite{Moresco2016} & $1.75$ & $202.0(40.0)$ &
\cite{Simon2005}\\\hline
$0.4497$ & $92.8(12.9)$ & \cite{Moresco2016} & $1.965$ & $186.0(50.4)$ &
\cite{Moresco2015}\\\hline
$0.47$ & $89.0(50.0)$ & \cite{Ratsimbazafy2017} &  &  & \\\hline
\end{tabular}

\end{center}

So we add to $\chi_{tot}^{2}$ the term%
\[
\chi_{H}^{2}=\sum_{i=1}^{31}\left(  \frac{H\left(  z_{i}\right)  -H_{i(obs)}%
}{\sigma_{H_{i}}}\right)  ^{2}\text{ .}%
\]

Using the expression $\chi_{tot}^{2}=\chi_{SN}^{2}+\chi_{CMB}^{2}+\chi
_{BAO}^{2}+\chi_{H}^{2}$, the likelihood function is given by%
\[
\mathcal{L}(\delta,n,\Omega_{\Psi i},\Omega_{bi})\propto exp[-\frac{\chi
_{tot}^{2}(\delta,n,\Omega_{\Psi i},\Omega_{bi})}{2}]\ \text{.}%
\]
So, by minimizing $\chi_{tot}^{2}$ (what is obviously equivalent to maximize
the likelihood function $\mathcal{L}$), we obtain the best fit values for the
parameters of the ITNADE and the IQNADE. For comparison, we also obtain the
best fit values for the noninteracting case (NADE) and for the $\Lambda CDM$
model. In the next section we show and discuss the results obtained.

\section{Results}

Table \textbf{5} below shows the individual best fits for all models
considered in this work. We integrate the equations of all models since the
redshift $z_{i}=3\times10^{5}$ - to increase $z_{i}$ in some orders of
magnitude did not affect the results. The $\chi_{\min}^{2}$, \textit{Akaike Information Criterion} (\textit{AIC}), \textit{Bayesian Information Criterion} (\textit{BIC}), $\Delta AIC$ and $\Delta BIC$ are also shown. Instead of showing the best fit values for the free parameters $\Omega_{\Psi i}$ and $\Omega_{bi}$, which are the relative densities of dark matter and baryon matter in the initial redshift $z_{i}$, their correspondig values today are shown.

\bigskip

\bigskip

\textbf{Table 5}: Values of model parameters of the ITNADE, IQNADE, NADE and
$\Lambda CDM$ from SNeIa, BAO, CMB and $H$. $\delta$ is dimensionless: for
ITNADE $\delta$ is in fact $\frac{\delta}{H_{0}}$, where $H_{0}=2.133h\times
10^{-39}MeV$ and $h=0.7$, and for IQNADE, $\delta$ is in fact $\delta M_{Pl}$,
where $M_{Pl}=2.436\times10^{21}MeV$ is the reduced Planck mass. $\Delta
AIC=AIC_{\operatorname{mod}el}-AIC_{\Lambda CDM}$ and $\Delta
BIC=BIC_{\operatorname{mod}el}-BIC_{\Lambda CDM}$.

\begin{center}
$%
\begin{tabular}
[c]{|p{1.5cm}|p{4cm}|p{4cm}|p{4cm}|p{4cm}|}\hline
& ITNADE & IQNADE & NADE & $\Lambda CDM$\\\hline
$n/\Omega_{\Lambda 0}$ & 2.445$_{-0.033}^{+0.033}$ & 2.453$_{-0.038}^{+0.038}$
& 2.405$_{-0.014}^{+0.014}$ & 0.6341$^{+0.0014}_{-0.0014}$\\\hline
$\Omega_{DM 0}$ & 0.2903$_{-0.0056}^{+0.0056}$ & 0.2885$_{-0.0059}^{+0.0059}$ & 0.2895$_{-0.0016}^{+0.0016}$ &
0.3150$_{-0.0014}^{+0.0014}$\\\hline
$\Omega_{b0}$ & 0.05542$_{-0.00078}^{+0.00078}$ & 0.05550$_{-0.00072}^{+0.00072}$ & 0.05740$_{-0.00027}^{+0.00027}$ &
0.05076$_{-0.00026}^{+0.00026}$\\\hline
$\delta$ & $-0.110_{-0.023-0.046-0.069}^{+0.023+0.046+0.069}$ &
$-0.065_{-0.014-0.028-0.042}^{+0.014+0.028+0.042}$ & - & -\\\hline
$\chi_{\min}^{2}$ & $1117.65$ & $1118.67$ & $1144.95$ & $1130.24$\\\hline
$AIC$ & $1125.65$ & $1126.67$ & $1150.95$ & $1136.24$\\\hline
$BIC$ & $1145.63$ & $1146.65$ & $1165.93$ & $1151.22$\\\hline
$\Delta AIC$ & $-10.59$ & $-9.57$ & $14.71$ & $-$\\\hline
$\Delta BIC$ & $-5.59$ & $-4.57$ & $14.71$ & $-$\\\hline
\end{tabular}
\ $

\end{center}

Note that the noninteracting model (NADE) has the same number of parameters
of the $\Lambda CDM$ ($n$, $\Omega_{\Psi i}$ and $\Omega_{bi}$ for the NADE
and $\Omega_{\Lambda i}$, $\Omega_{\Psi i}$ and $\Omega_{bi}$ for the $\Lambda
CDM$). For a large number of degrees of freedom, the distribution of $\chi
^{2}$ is gaussian, with mean equal to the number of degrees of freedom,
$\chi^{2}=\upsilon=1091-3=1088$ in this case, and standard deviation
$\sigma=\sqrt{2\upsilon}=46.7$. If we define the criterion that values of
$\chi_{\min}^{2}$ whithin an interval of $2\sigma$ around the best value
$\chi^{2}=\upsilon$ are acceptable, or in other words, if we define the
criterion that fits whose $\chi_{\min}^{2}$ are whithin the interval
$994.7<\chi^{2}<1181.3$ are acceptable, then by this $\chi^{2}$ criterium, we
can say that the NADE model fits well the present set of observational data
(in fact, for the NADE $\chi_{\min}^{2}=1144.95\simeq\upsilon+1.2\sigma$, and
for the $\Lambda CDM$ $\chi_{\min}^{2}\simeq1130.24=\upsilon+0.9\sigma$). For
more details about the $\chi^{2}$ criterion see, e. g. \cite{livrobarlow}.

We can see by the values of $\chi_{\min}^{2}$ showed in the table 5, that the
two interacting models fits the data better than the $\Lambda CDM$. But this
improvement on the fit is sufficient to justify the introduction of one more
free parameter (the coupling constant $\delta$) in the NADE model? This
question can be answered using, for example, the \textit{AIC} \cite{Akaike}
and \textit{BIC} \cite{BIC} criteria. We can use the \textit{AIC} and
\textit{BIC} criteria to answer if a given model is prefered by the data or
not, or in other words, if the data furnishes sufficient evidence in favor of
a given model. Obviously, what we want to know in this work is if there exists
evidence in favor of an interaction between dark energy and dark matter.

The \textit{AIC} is basically a frequentist criterion, and for a large set of
data and Gaussian errors, it is given by%
\begin{equation}
AIC=-2\ln\mathcal{L}_{\max}+2p\text{ ,} \label{AIC}%
\end{equation}
where $p$ is the number of free parameters of the model. If we want to know if
there exists evidence in favor of a given model, say \textit{model 1}, in
relation to another model, \textit{model 2}, we need to compute $\Delta
AIC=AIC_{\operatorname{mod}el\text{ 1}}-$ $AIC_{\operatorname{mod}el\text{ 2}%
}$. If $4<\Delta AIC<7$ there is evidence in favor of the \textit{model 2},
that is, the model with minor \textit{AIC} value. If $\Delta AIC>10$ such an
evidence is strong. For detailed discussions about \textit{AIC} and
\textit{BIC} criteria see, e. g., \cite{livroburnham} and \cite{livrohastie}.

The \textit{BIC} follows from a Gaussian approximation to the Bayesian
evidence in the limit of large sample size \cite{Trotta}:%
\begin{equation}
BIC=-2\ln\mathcal{L}_{\max}+p\ln N\text{ ,} \label{BIC}%
\end{equation}
where $p$ is the number of free parameters and $N$ is the number of data
points. In the same manner as for AIC, if $2<\Delta BIC<6$, there is positive
evidence in favor of the model with minor \textit{BIC} value. Again, if
$\Delta BIC>10$, such an evidence is strong.

From table \textbf{5}, we see that the \textit{AIC} and \textit{BIC} criteria
furnishes strong evidence against the noninteracting case (NADE), in relation
to the $\Lambda CDM$ model. Such a conclusion has already been obtained in
\cite{testenade1} and \cite{testenade2}, from different data sets. However,
for both interacting models, ITNADE and IQNADE, \textit{AIC} criterion
furnishes strong evidence in favor of the interacting models, whereas
\textit{BIC} criterion furnishes moderate evidence. Therefore, considering
both the criteria, in the present work \textit{we have obtained strong
evidence in favor of both the interacting models.} These result, combined with
the fact that for both ITNADE and IQNADE the coupling constant is nonvanishing
at more than $3\sigma$ confidence level, \textit{give us significant
evidence of an interaction between dark energy and dark matter. }Furthermore,
the sign of the coupling is compatible with dark energy decaying into dark
matter, alleviating the coincidence problem.

\begin{figure}[!hbp]
\begin{center}
\includegraphics[width=17.54cm,height=6.81cm]{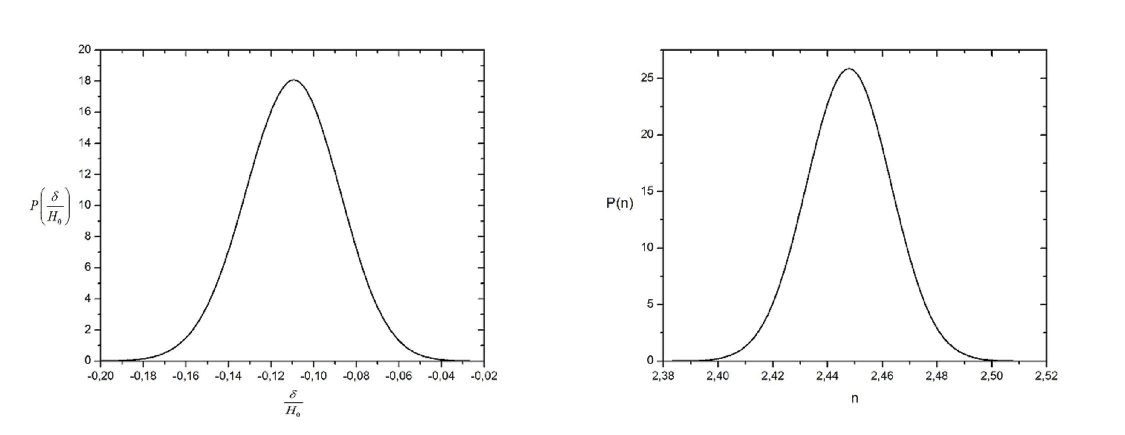}
\end{center}
\caption{Marginalized probability distribuctions of $\frac{\delta}{H_{0}}$ and $n$ for the ITNADE model.}%
\label{distributions_tac}%
\end{figure}

\begin{figure}[!hbp]
\begin{center}
\includegraphics[width=17.54cm,height=6.81cm]{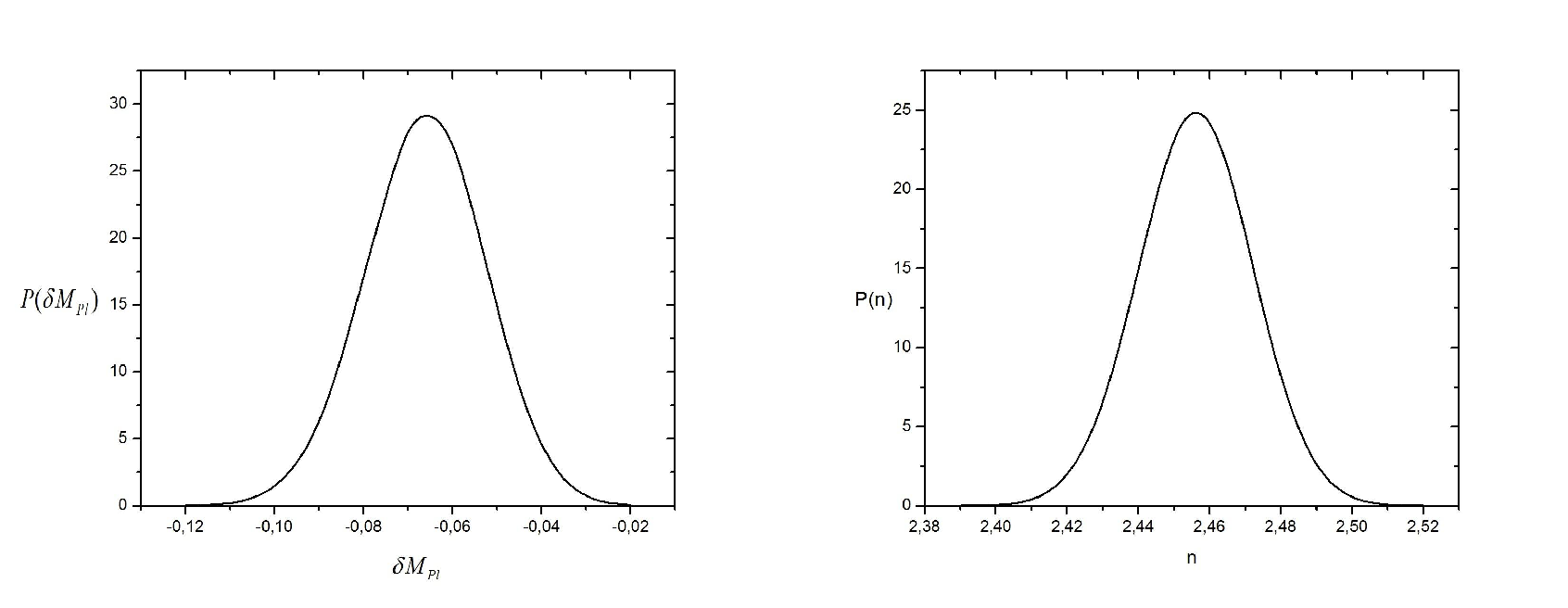}
\end{center}
\caption{Marginalized probability distribuctions of $\delta M_{Pl}$ and $n$ for the IQNADE model.}%
\label{distributions_quint}%
\end{figure}

\begin{figure}[!hbp]
\includegraphics[width=17.00cm,height=5.02cm]{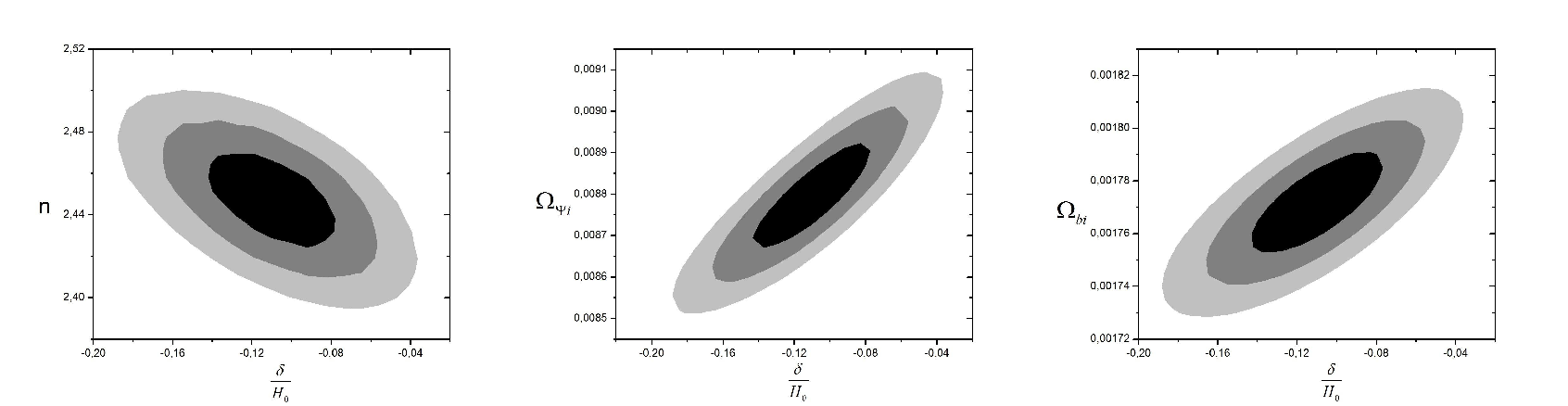}
\caption{Confidence regions of $1\sigma $, $2\sigma $ and $3\sigma $ for two parameters for the ITNADE model.}%
\label{bidimensionals_tac}%
\end{figure}

\begin{figure}[!hbp]
\includegraphics[width=17.00cm,height=5.02cm]{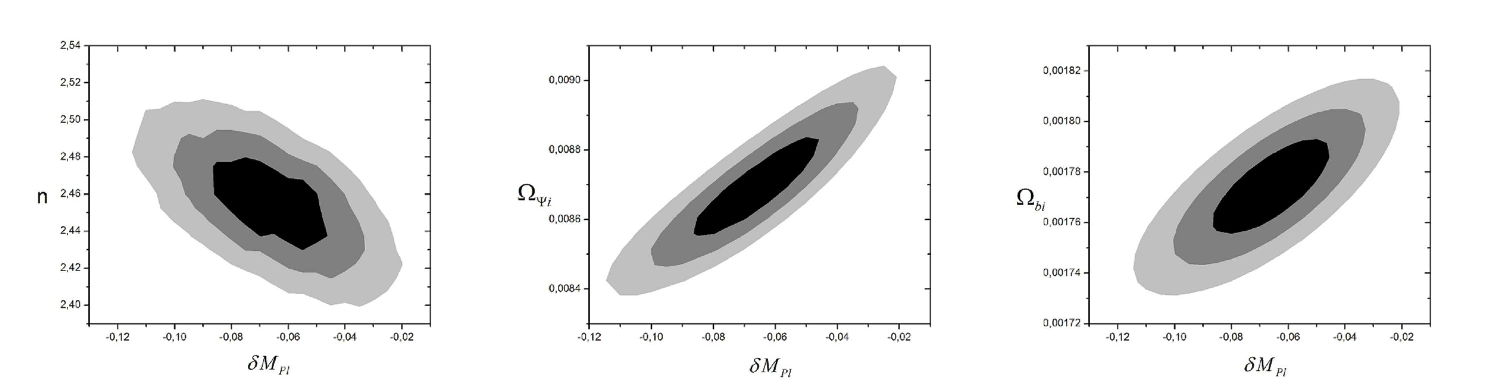}
\caption{Confidence regions of $1\sigma $, $2\sigma $ and $3\sigma $ for two parameters for the IQNADE model.}%
\label{bidimensionals_quint}%
\end{figure}

Figures \ref{distributions_tac} and \ref{distributions_quint} show the marginalized probability
distribuctions of the (dimensionless) coupling constant $\delta$ and $n$,
whereas figures \ref{bidimensionals_tac} and \ref{bidimensionals_quint} show the two parameter confidence
regions of $1\sigma$, $2\sigma$ and $3\sigma$ for the ITNADE and the IQNADE models.

It is interesting to note that there is a little degeneracy between the
coupling constant $\delta$ and $n$, so that in both interacting models $n$ is
bigger than in the noninteracting case. The differences, however, are less
than 1$\sigma $.

In summary, we have derived two field theory models of interacting dark energy
and have made the comparison of these models with recent observational data.
We have also made the comparison of the noninteracting and $\Lambda CDM$
models with the data. From the aplication of the \textit{AIC} and \textit{BIC}
model selection criteria, \textit{we have obtained strong evidence in favor of
the two interacting models}. Moreover, the coupling constants of the two
models are nonvanishing at more than $3\sigma$ confidence level. Therefore,
\textit{we have obtained significant evidence of an interaction in the dark
sector of the universe}. This conclusion goes in the same direction of other
works in recent years, e. g. \cite{interacaoelcio}, \cite{sandro} - \cite{interacaoevidencia}.

\end{document}